# A Simple Route towards High-Concentration Surfactant-Free Graphene Dispersions

Jiantong Li[*], Fei Ye, Sam Vaziri, Mamoun Muhammed, Max C. Lemme and Mikael Östling

KTH Royal Institute of Technology, School of Information and Communication Technology, Electrum 229, 16440 Kista, Sweden

**Abstract** A simple solvent exchange method is introduced to prepare high-concentration and surfactant-free graphene liquid dispersion. Natural graphite flakes are first exfoliated into graphene in dimethylformamide (DMF). DMF is then exchanged by terpineol through distillation, relying on their large difference in boiling points. Graphene can then be concentrated thanks to the volume difference between DMF and terpineol. The concentrated graphene dispersions are used to fabricate transparent conductive thin films, which possess comparable properties to those prepared by more complex methods.

[*]Corresponding author. Fax: +46 8 752 7850. E-mail address: jiantong@kth.se (J. Li)

Graphene has attracted huge interest because of its outstanding electronic, mechanical and thermal properties [1]. Recently, researches on large-scale and high-throughput methods for mass production of graphene have come into focus to address future applicability of graphene on an industrial scale. An efficient method towards mass production is the direct exfoliation of graphite into single-layer or few-layer graphene in certain solvents. It has been shown that dimethylformamide (DMF) [2] and N-methyl-2-pyrrolidone [3] are ideal organic solvents for the exfoliation of high-quality monolayer graphene. However, one critical drawback of using these solvents is that the produced graphene dispersion is at relatively low concentrations, typically < 0.01 mg/mL. Recently, significant progress has been made to increase the graphene concentration to over 0.1 mg/mL through techniques such as spontaneous exfoliation in superacids [4], direct exfoliation assisted by long-time (several weeks) sonication [5], and post concentration based on solvent exchange [6,7]. However, superacids with their high reactivity may not be suitable for further processing [7], and long-time sonication may not prove efficient for mass production. Meanwhile, the demonstrated solvent exchange techniques [6,7] involve more complicated processes, waste of exfoliated graphene, or require stabilizing polymers. In this work, we demonstrate a novel but simple solvent exchange technique to prepare high-concentration graphene dispersions. The essence of our technique is that we use two solvents with significantly different boiling points, so that the solvent exchange can be readily accomplished through a distillation process.

Our solvent exchange technique is illustrated in Fig. 1. First, a mixture of ~ 2 mg/ml natural graphite flake (Sigma-Aldrich, product no. 332461) in DMF was sonicated (Branson 2510E-MTH bath ultrasonicator) for ~ 20 h. The resultant suspension was centrifuged at > 10000 rpm for 15 min to sediment thick flakes and the supernatant was harvested (Fig. 1a). To ensure the removal of thick flakes, the centrifugation was repeated once. Then terpineol was added to the harvested supernatant (Fig. 1b). A vacuum distillation process was used where the pressure is reduced and the distillation commences at a lower temperature than the solvents' boiling points. The graphene dispersion in the mixed DMF/terpineol solvent was heated to 80 °C by a water bath. When the pressure was reduced to ~30 mbar, DMF began to evaporate whereas terpineol was stable until the pressure reached values below 10 mbar. After DMF was boiled off, the remaining graphene/terpineol dispersion was harvested (Fig. 1c). We observed some black particles suspended in this harvested dispersion, but a simple shaking and a rough sonication for only a few minutes effectively redispersed all suspended particles and pure (particle-free) graphene dispersion was obtained (Fig. 1d). We choose terpineol as the second solvent for three reasons. First, its boiling point (~219 °C) is significantly higher than that of DMF (~153 °C), which is desired for distillation. Second, terpineol has been demonstrated to be an excellent solvent for stable graphene dispersions [7]. Last and most importantly, terpineol-based solutions/inks play an important role in the present ink-jet printing technology [8], which is promising for the emerging printed and flexible electronics.

Since almost all the exfoliated graphene is transferred into the final terpineol dispersion, the graphene concentration can be well controlled by the volume ratio ($R_V$) of DMF to terpineol. We have managed to prepare graphene dispersions with $R_V$ up to ~ 40. Graphene dispersions from $R_V$ = 40 can be stable (i.e., without any occurrence of sediments or visible particles) for about 10 h. And those from $R_V$ = 20 can be stable for ~ 30 h. Based on the measured optical absorbance [3], the highest graphene concentration (from $R_V$ = 40) was estimated to be about 0.39 mg/mL. Since these graphene dispersions were prepared without surfactant/polymer stabilization or long-time sonication, the attained value of 0.39 mg/mL is quite encouraging. In addition, $R_V$ = 40 does not constitute an upper limit for this process. The maximum $R_V$ is expected to be limited only by the graphene solubility in terpineol.

We have used transmission electron microscopy (TEM, JEOL JEM 2100F) to observe the flake edges and provide an accurate way to measure the layer number of the flakes (as first shown in [9]). From TEM images, we randomly chose and analyzed 25 flakes for each graphene sample before (Fig. 2a and S1) and after distillation (Fig. 2b and S2). It was found that before distillation, the flakes mostly consist of 4~5 graphene layers (Fig. 2c), while after distillation, they mostly comprise 4~7 layers (Fig. 2d). It suggests that the graphene agglomeration during the distillation process is not severe.

The graphene dispersion was used to fabricate transparent conductive thin film by vacuum filtration [10]. Fig. 3a shows such films on glass slides. In Fig. 3b, the scanning electron micrograph (SEM, Zeiss Ultra 55) image clearly shows dense and uniform graphene flakes (with most lateral dimensions at the level of 100 nm) throughout the film. Fig. 3c displays the Raman spectra of original graphite flake, and graphene before and after distillation. The intensity ratio of the D and G bands is ~ 0.5 for graphene before distillation, and is ~ 0.65 for graphene after distillation, which indicates that the distillation process does not severely increase defects [3,7,10]. Fig. 3d plots the optical transmittance versus the sheet resistance for the fabricated transparent conductive graphene thin films. Their performance (for example, sheet resistance ~ 6 k$\Omega$/□ at transmittance ~ 60% for wavelength $\lambda$ = 550 nm) is comparable with those produced directly from surfactant/polymer-stabilized graphene liquid dispersions [7,10], though still inferior to those based on the growth [11] or reduction [12] methods.

In summary, we propose a simple yet general route to prepare high-concentration surfactant-free graphene liquid dispersions based on a distillation-assisted solvent exchange technique. Our experiments demonstrate that the prepared graphene dispersions contain few-layer graphene flakes without severe defects. The dispersions reach graphene concentrations as high as 0.39 mg/mL and are stable for at least 10 h without any surfactant/polymer stabilization. Despite the simplicity of our technique, the graphene dispersions can be used to fabricate transparent conductive thin films, which have comparable optical and electrical properties with those from other more complicated liquid methods.

**Acknowledgements.** The authors would like to thank Anand Srinivasan and Reza Sanatinia for fruitful discussions regarding Raman spectroscopy, and Mazher Ahmed Yar for SEM measurements. Support and sponsorship by the European Research Council through the Advanced Investigator Grant (OSIRIS, No. 228229), by the European Commission through FP7 Project NanoMmune (NMP4-SL-2008-214281) and by a seed project (Call 3, 2011) from VINN Excellence iPack Center are gratefully acknowledged.

**Appendix A. Supplementary data.** Supplementary material associated with this article can be found in the online version.

**References**

[1] Geim AK, Novoselov KS. The Rise of Graphene. Nat Mater 2007;6(3):183-91.

[2] Blake P, Brimicombe PD, Nair RR, Booth TJ, Jiang D, Schedin F, et al. Graphene-Based Liquid Crystal Device. Nano Lett 2008;8(6):1704-8.

[3] Hernandez Y, Nicolosi V, Lotya M, Blighe FM, Sun Z, De S, et al. High-Yield Production of Graphene by Liquid-Phase Exfoliation of Graphite. Nat Nanotechnol 2008;3(9):563-8.

[4] Behabtu N, Lomeda JR, Green MJ, Higginbotham AL, Sinitskii A, Kosynkin DV, et al. Spontaneous High-Concentration Dispersions and Liquid Crystals of Graphene. Nat Nanotechnol 2010;5(6):406-11.

[5] Khan U, O'Neill A, Lotya M, De S, Coleman JN. High-Concentration Solvent Exfoliation of Graphene. Small 2010;6(7):864-71.

[6] Zhang X, Coleman AC, Katsonis N, Browne WR, Wees BJ, Feringa BL. Dispersion of Graphene in Ethanol Using a Simple Solvent Exchange Method. Chem Commun 2010; 46(40):7539-41.

[7] Liang YT, Hersam MC. Highly Concentrated Graphene Solutions via Polymer Enhanced Solvent Exfoliation and Iterative Solvent Exchange. J Am Chem Soc 2010;132(50):17661-3.

[8] Lee HH, Chou KS, Huang KC. Inkjet Printing of Nanosized Silver Colloids. Nanotechnology 2005;16(10):2436-41.

[9] Meyer JC, Geim AK, Katsnelson MI, Novoselov KS, Booth TJ, Roth S. The structure of suspended graphene sheets. Nature 2007; 446(7131):60-3.

[10] Green AA, Hersam MC. Solution Phase Production of Graphene with Controlled Thickness via Density Differentiation. Nano Lett 2009;9(12):4031-6.

[11] Kim KS, Zhao Y, Jang H, Lee SY, Kim JM, Kim KS, et al. Large-scale pattern growth of graphene films for stretchable transparent electrodes. Nature 2009;457(7230):706-10.

[12] Becerril HA, Mao J, Liu Z, Stoltenberg RM, Bao Z, Chen Y. Evaluation of solution-processed reduced graphene oxide films as transparent conductors. ACS Nano 2008;2(3):463-70.

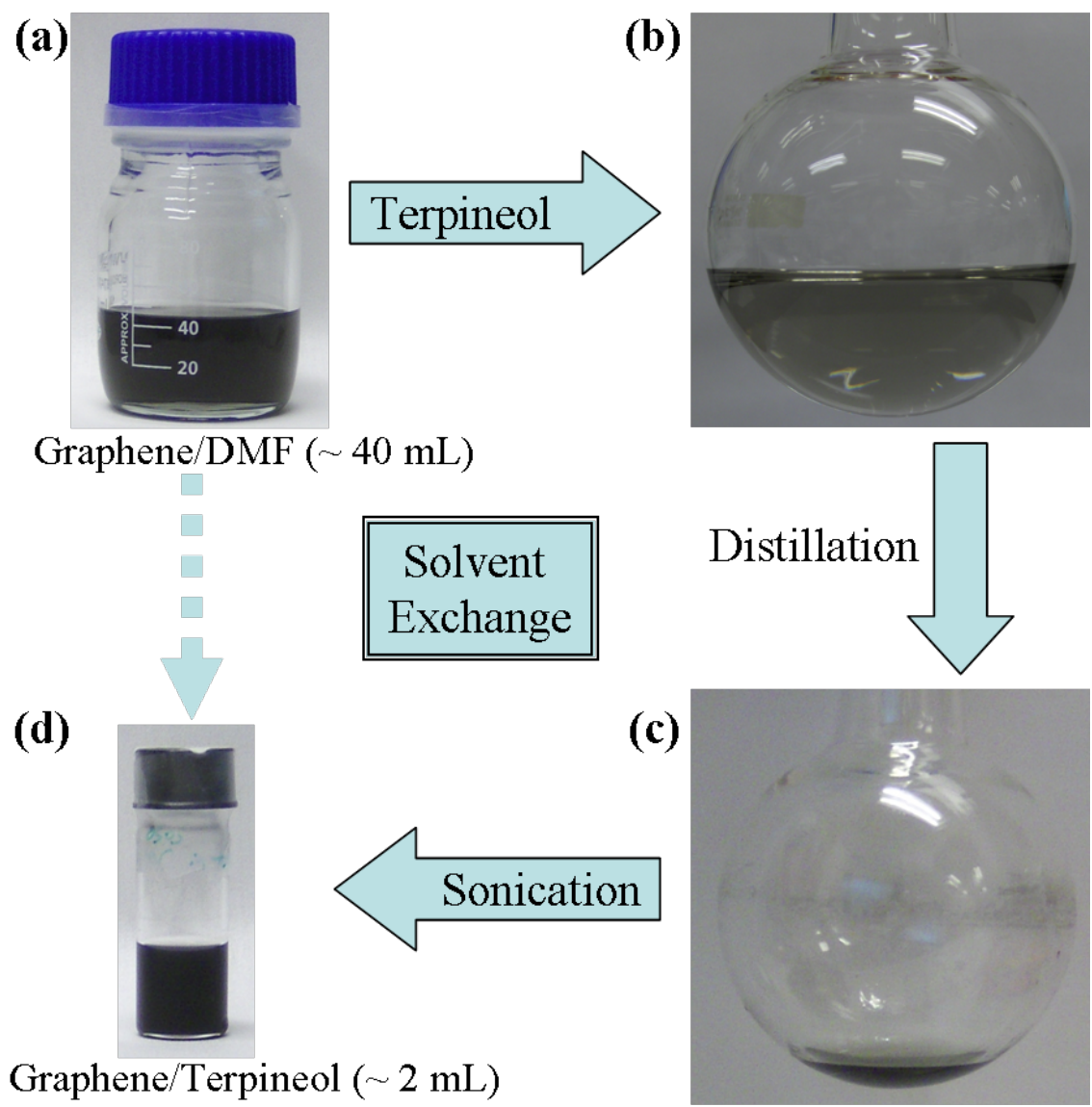

**Fig. 1.** Illustration of the distillation-assisted solvent exchange technique ($R_V$=20) proposed in this work. (a) Exfoliated graphene in DMF. (b) Graphene/DMF dispersion mixed with terpineol. (c) Graphene/terpineol dispersion obtained from distillation of (b). (d) Pure (particle-free) graphene/terpineol dispersion obtained from a rough sonication of (c).

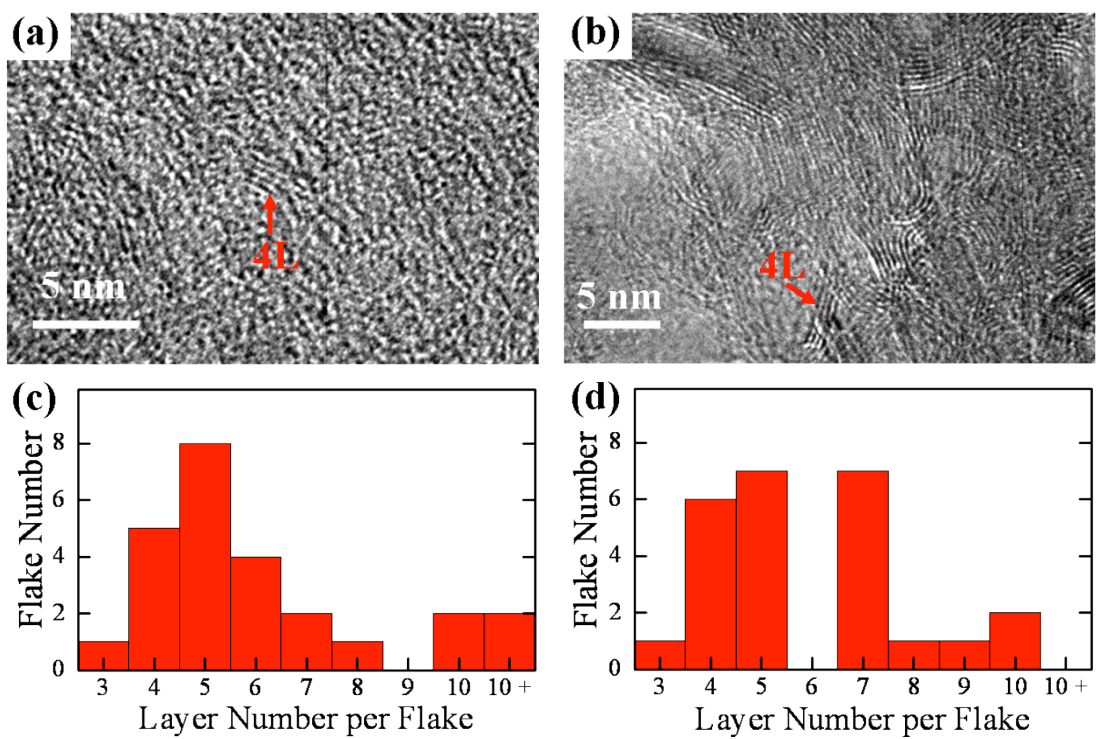

**Fig. 2**. TEM characterization of graphene films. (a-b) High magnification TEM images showing the flake edges of (a) graphene from DMF dispersion (before distillation) and (b) graphene from terpineol dispersion (after distillation). As examples, the arrows indicate flakes consisting of 4 graphene layers. For TEM samples in (b), the concentrated graphene dispersion ($R_V$=40) was diluted 20 times with terpineol. (c-d) Histograms of the layer number per flake for graphene before distillation (c) and after distillation (d).

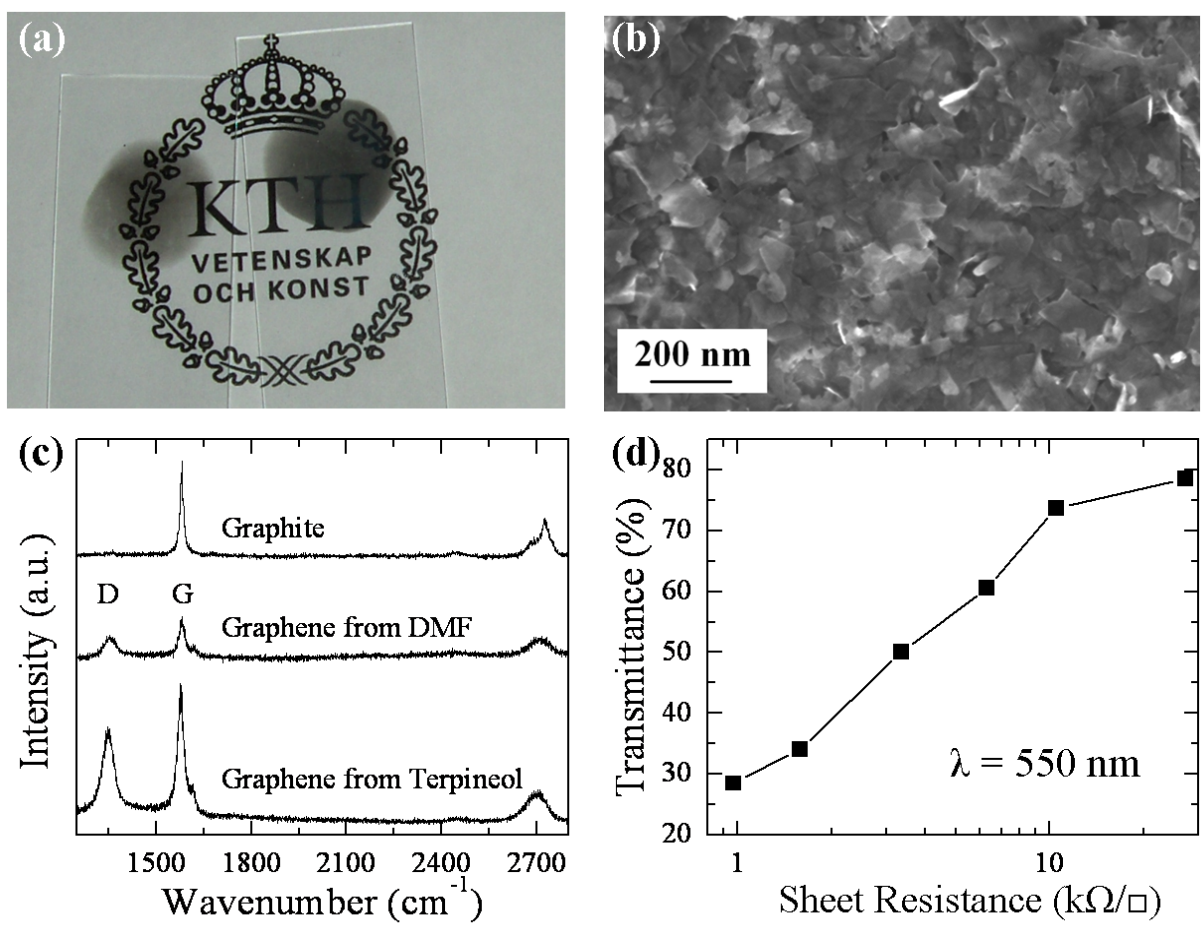

**Fig. 3**. Characterizations of graphene thin films fabricated by vacuum filtration. (a) Photographs of two conductive graphene thin films on glass slides. The glass slides are 26 mm wide. In this picture, the glass slides were put on a paper with KTH logo to demonstrate the transparence of the graphene thin films. (b) An SEM image of one conductive graphene thin film. (c) Raman spectrum measured at 514 nm excitation for a graphene thin film fabricated by vacuum filtration of the terpineol dispersion (after distillation). For comparison, it also shows Raman spectra for the original graphite flake and a graphene film fabricated by drop casting of the DMF dispersion (before distillation). (d) Optical transmittance ($\lambda$ = 550 nm) versus sheet resistance for the graphene thin films.